\documentclass{optica-article}

\journal{opticajournal} 

\articletype{Research Article}

\usepackage{lineno}

\begin{document}

\title{Thermo-optic modulator with ultra-high extinction ratio for low-loss silicon nitride integrated photonics
}

\author{
	Dmitriy Serkin,\authormark{1,3} Kirill Buzaverov,\authormark{1,2,3} Aleksandr Baburin,\authormark{1,2} Evgeny Sergeev,\authormark{1} Sergey Avdeev,\authormark{1} Evgeniy Lotkov,\authormark{1} Sergey Bukatin,\authormark{1} Ilya Stepanov,\authormark{1} Aleksey Kramarenko,\authormark{1} Ali Amiraslanov,\authormark{1} Ilya Ryzhikov,\authormark{1} and Ilya Rodionov\authormark{1,2,*}}

\address{\authormark{1}Shukhov Labs, Quantum Park, Bauman Moscow State Technical University, 105005 Moscow, Russia\\
\authormark{2}Dukhov Automatics Research Institute (VNIIA), Moscow 127055, Russia\\
\authormark{3}Authors contributed equally\\
}

\email{\authormark{*}irodionov@bmstu.ru} 


\begin{abstract*}
	Extremely low-loss silicon nitride integrated circuits is a potential platform for a growing number of frontier applications in quantum technologies, high-performance and analog computing, nonlinear optics, light detection and ranging (LiDAR), and biotechnologies. However, eﬃcient optical modulation with a wide frequency response, high contrast, low power and scalable manufacturing remains one of the key challenges for silicon nitride integrated photonics. Here, we propose an integrated thermo-optic phase shifter with isolation trenches operating in the C-band. The fabricated thermo-optic modulator capable to achieve a $\pi$-phase shift at a power consumption of 65~mW, bandwidth of 12~kHz, and extinction ratio (ER) over 80~dB. Moreover, we systematically demonstrate its compatibility with low-loss silicon nitride photonic integrated circuits with microring resonators exibiting an average quality factor more than $5.9\times 10^6$, which correspond to propagation loss of 0.058~dB/cm.
\end{abstract*}

\section{Introduction}

The field of integrated photonics has been developing for decades and is crucial to a variety of emerging technologies. These include large bandwidth coherent transceivers for datacenters, high throughput network switches for next-generation communication systems, photonic neural and quantum computing systems, solid-state LiDAR devices for automotive and agricultural applications as well as critical infrastructure security, high-precision portable devices for non-invasive analysis, biosensing, and spectroscopy \cite{10.1038/s41467-024-44750-0, 10.1002/lpor.202400508}. Intense research has resulted in a variety of photonic platforms. Among these, silicon (SOI), silicon nitride, III-V semiconductors (InP, GaAs), and lithium niobate (LNOI) have achieved a high level of maturity. Silicon nitride is a unique material for integrated photonics due to its broad transmission spectrum, covering wavelengths from the visible to the mid-infrared, its ultralow optical loss, and its full CMOS compatibility \cite{10.1038/s41566-021-00903-x, 10.1038/s41467-021-21973-z, 10.1126/science.abh2076}. However, because it has  a direct energy bandgap, silicon nitride is not suitable for fabricating  lasers, modulators, or photodetectors \cite{10.1038/s41467-021-26804-9}. Progress in heterogeneous integration of silicon nitride with SOI, III-V gain materials, and LNOI platforms \cite{10.1364/PRJ.452936, 10.1038/s41467-023-39047-7} has led to demonstrations of highly coherent integrated lasers with ultra-narrow linewidth \cite{10.1038/s41467-021-26804-9, 10.1364/OL.433636}, high-efficiency photodetectors \cite{10.1038/s41377-022-00784-x, 10.1038/s41467-022-34100-3}, and high-bandwidth modulators \cite{10.1038/s41467-018-05846-6, 10.1038/s41563-018-0208-0}.

Optical modulators play a key role in controlling the amplitude and phase of optical signals. The fundamental element of an optical modulator is the phase shifter, which provides precise control over the phase delay of light propagating through the waveguide. This ability to dynamically tune the phase underlies the operation of interferometric structures, delay lines, switches, and other components required for information processing and other applications. 

The development of efficient phase shifters for silicon nitride photonics faces significant technological challenges. One of the main difficulties is reducing optical loss and power consumption. Various technologies offer potential solutions, but they often involve trade-offs. 

Phase shifters based on microelectromechanical systems (MEMS) demonstrate excellent energy efficiency, consuming only $\sim 100$~nW \cite{10.1364/OL.40.003556}, but they require high driving voltages and complex fabrication, which limit their use \cite{10.1038/s41467-024-44750-0a}. Ferroelectric phase shifters achieve ultrafast modulation (500~GHz \cite{10.1063/1.5086868}), but suffer from high insertion loss (IL) and integration challenges \cite{10.1002/lpor.202400624}. Plasma‑dispersion phase shifters in silicon are compact and fast, but they intrinsically suffer from free‑carrier absorption, thermal drift, and phase‑to‑amplitude crosstalk, and the mechanism is not directly transferable to Si$_3$N$_4$ \cite{10.1038/s41467-024-45130-4, 10.1117/1.AP.3.2.024003}. Transparent conducting oxides (e.g., indium tin oxide) exploit free-carrier dispersion in the epsilon-near-zero regime to achieve sub-nanosecond phase shifts via electrostatic carrier modulation \cite{10.48550/arXiv.2412.19306, 10.1038/s41598-022-09973-5, 10.1364/AOP.448391}, though insertion loss above 5~dB require further material optimization. Phase-change materials (PCM) such as Sb$_2$Se$_3$ on silicon exhibit low insertion loss ($\sim 0.36$~dB) and a high phase modulation efficiency (0.09$\pi$/\textmu m \cite{10.1186/s43074-022-00070-4}), but their limited cyclability ($\sim$10$^4$ switching cycles \cite{10.1002/adfm.202304601}) restricts their use to applications with moderate switching requirements.

\begin{figure}[t]
	\centering
	\includegraphics[width=\linewidth]{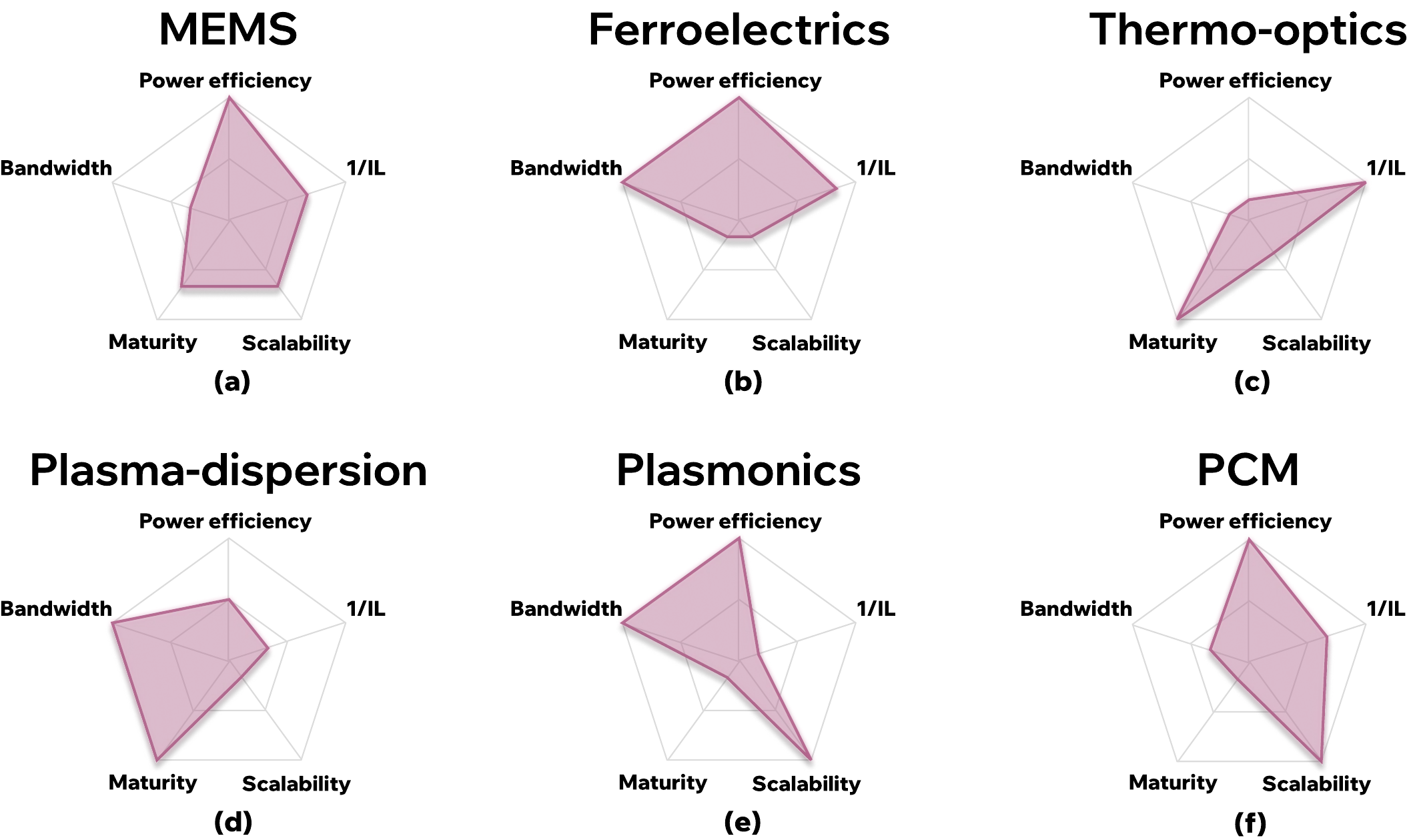}
	\caption{Comparison of the main characteristics of various phase shifting technologies: (a) MEMS; (b) Ferroelectrics; (c) Thermo-optics; (d) Plasma-dispersion; (e) Plasmonics; (f) Phase-change materials.\label{TOPS:Fig1}}
\end{figure}

In light of these trade-offs, thermo-optic phase control offers a favorable balance of fabrication simplicity, optical performance, and reliability for silicon nitride platform. Table \ref{TOPS:Table1} presents recent advances in thermo-optic phase shifters based on silicon nitride PICs. A comparison of phase shifting approaches is presented in Figure \ref{TOPS:Fig1}.

\begin{table}[b] 
	\caption{Comparison of the thermo-optic phase shifters based on silicon nitride photonic integrated circuits for the C-band.\label{TOPS:Table1}}
	\centering
	\begin{tabularx}{1\textwidth}{ >{\centering\arraybackslash}X | >{\centering\arraybackslash}X | >{\centering\arraybackslash}X | >{\centering\arraybackslash}X | >{\centering\arraybackslash}X | >{\centering\arraybackslash}X}
		\hline
		\textbf{Heater structure} & \textbf{Length, \textmu m} & \textbf{$P_\pi$, mW} & \textbf{$T_\text{rise}$, \textmu s} & \textbf{Modulator ER, dB} & \textbf{Ref.}\\
		\hline

		Ni-Cr, single-strip     & 800         & 50   & 50    & 20   & \cite{10.1364/OE.396969} \\
		Al, suspended           & 1000        & 100  & 26.43 & ---  & \cite{10.3390/photonics8110496} \\
		Au, PT-symmetry braking & 1000        & 157  & 8.5   & 7    & \cite{10.48550/ARXIV.2408.15139} \\
		Ni-Cr, suspended        & 300         & 12   & 1300  & 29.7 & \cite{10.1364/PRJ.507548} \\
		Cr-Au, single-strip     & 2000        & 42.7 & 110   & 20   & \cite{10.1364/OE.547586} \\
		Pt, single-strip        & 1000        & 385  & ---   & ---  & \cite{10.1088/2633-4356/ac168c} \\
		Cr-Au, suspended        & $\sim$ 1000 & 290  & ---   & 30   & \cite{10.1364/OE.416053} \\

		Ti, suspended           & 400         & 65   & 35    & 80   & This work \\ 

		\hline
	\end{tabularx}
\end{table}

\begin{figure}[t]
	\centering
	\includegraphics[width=\linewidth]{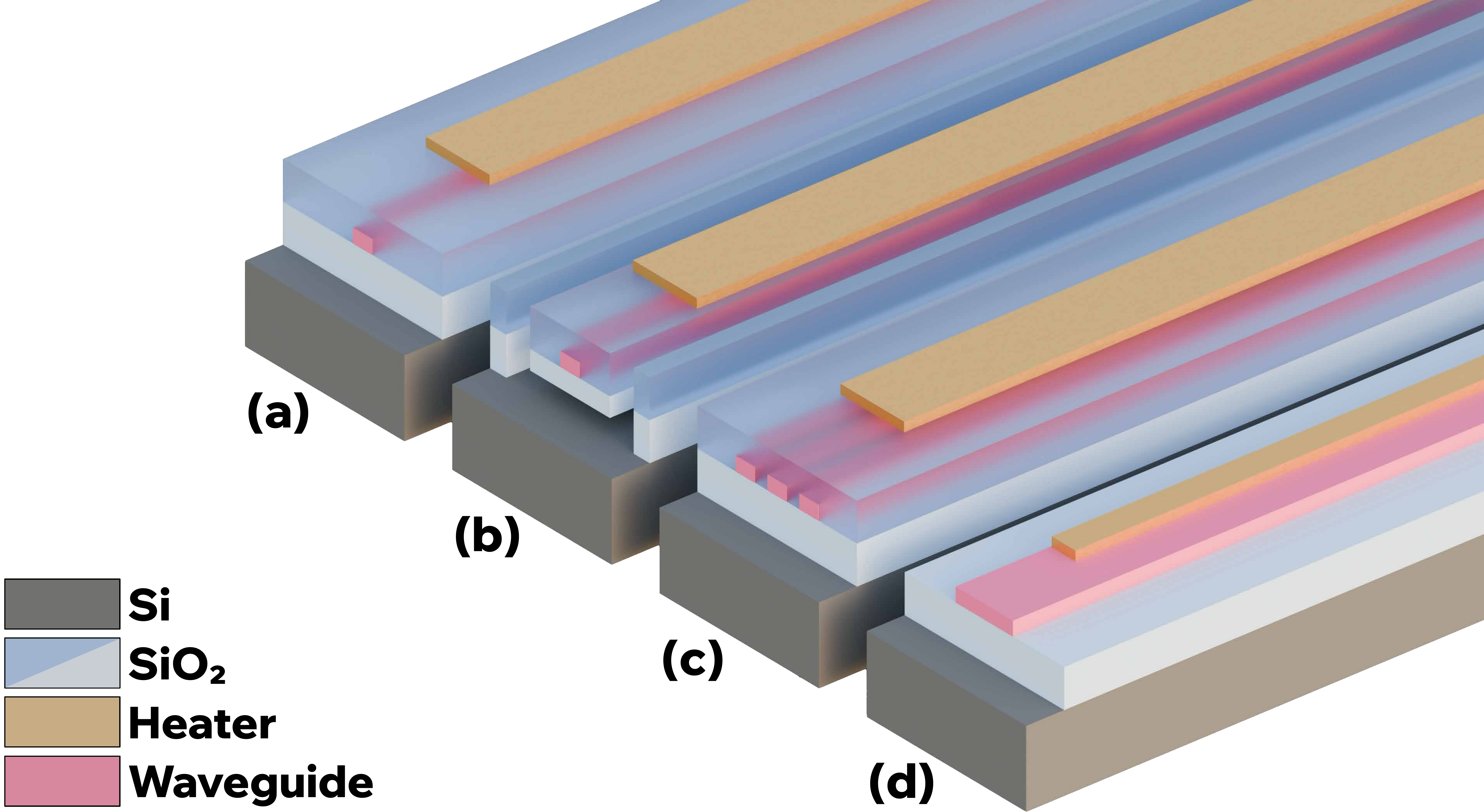}
	\caption{Cross sections of various types of thermo-optic phase shifters: (a) Single-strip phase shifter; (b) Suspended single-strip phase shifter; (c) Phase shifter with folded waveguide; (d) PT-symmetry breaking phase shifter.\label{TOPS:Fig2}}
\end{figure}

The basic design principle of thermo-optic phase shifters involves use of a direct waveguide with a parallel heating element, as illustrated in Figure \ref{TOPS:Fig2}(a). This design ensures stable operation regardless of the direction of optical signal propagation. The heater is made of electrically conductive material, which passes an electric current and is therefore subjected to Joule heating.Various metals or their compounds, such as platinum \cite{10.1364/OL.469358}, titanium nitride \cite{10.1364/OE.27.005851a}, tungsten \cite{10.1364/OL.467779}, gold \cite{10.48550/ARXIV.2408.15139}, indium tin oxide \cite{10.1364/OE.386959}, etc., are used to manufacture heaters. Due to the relatively low thermo-optic coefficient of Si$_3$N$_4$ of $2.51\times10^{-5}$~K$^{-1}$ \cite{10.1364/OE.477102}, conventional thermo-optic phase shifters have a power consumption of more than 100~mW and a frequency bandwidth in order of kHz \cite{10.1088/2633-4356/ac168c}.

There are several ways to improve the key features of thermo-optic phase shifters, such as power consumption and response time. The most straightforward approaches include the use of materials with low thermal conductivity, placing the heating element closer to the waveguide, and reducing the heat capacity of the waveguide by shortening its length. Each of these methods, however, has significant drawbacks. Heating elements with a low heat transfer capacity do not support high switching speeds, reducing the distance between the waveguide and the heater leads to an increase in optical loss. Additionally, shortening the waveguide causes the heater temperature to rise above critical values during operation, which results in phase shifting errors and heater failure. Novel approaches to optimizing the design of phase shifters include the fabrication of suspended region via deep trenches in the cladding and substrate near the heater \cite{10.3390/mi12050534, 10.3390/mi13111925}, the use of folded \cite{10.1364/OE.559081} or multimode waveguides \cite{10.1364/OPTICA.7.000003}, as well as the use of the effect of parity-time symmetry breaking \cite{10.1364/OE.567666}, integration of doped silicon heaters \cite{10.3390/photonics8110496} or polymer waveguides \cite{10.1364/OE.547586}.

Despite these advances, current approaches do not provide phase shifters with low power consumption, high switching speed, low insertion loss and a mature fabrication process. In this paper, we propose an integrated thermo-optic phase shifter for silicon nitride photonics based on a single strip titanium heater with isolation trenches. We carefully optimized the phase shifter design for the C-band using thermal and electromagnetic simulations. This allowed us to identify ways to minimize power consumption on our integrated photonic platform, as described in Section \ref{TOPS:Sec31}. We fabricated  several series of unbalanced Mach-Zehnder interferometers (MZIs) with optimized phase shifters for optical and electrical characterization of both static and dynamic properties. We demonstrate a $\pi$-phase shift power consumption of 195~mW, which is reduced down to 65~mW by introducing isolation trenches in the SiO$_2$ cladding. Both types of phase shifters exhibit a bandwidth of $12$~kHz ($-3$ dB cutoff) with 10-90\% rise and fall times of less than 35~\textmu s. For both designs, we measured an extinction ratio ER of over 80~dB in MZI-based amplitude modulators. Our results illustrate the potential of efficient metal-based phase shifters in mass-produced ultralow-loss silicon nitride photonic integrated circuits. We achieved microresonator quality (Q) factors over $9.6\times10^6$ (propagation loss $\sim0.033$~dB/cm), enabling their use in a variety of systems that do not require modulators operating at microwave frequencies, such as programmable notch filters and optical phased arrays \cite{10.1364/OE.416053, 10.1126/sciadv.abq2196, 10.3390/app9020255a}.

\section{Thermo-optic phase shifter design}

\begin{figure}[t]
	\centering
	\includegraphics[width=\linewidth]{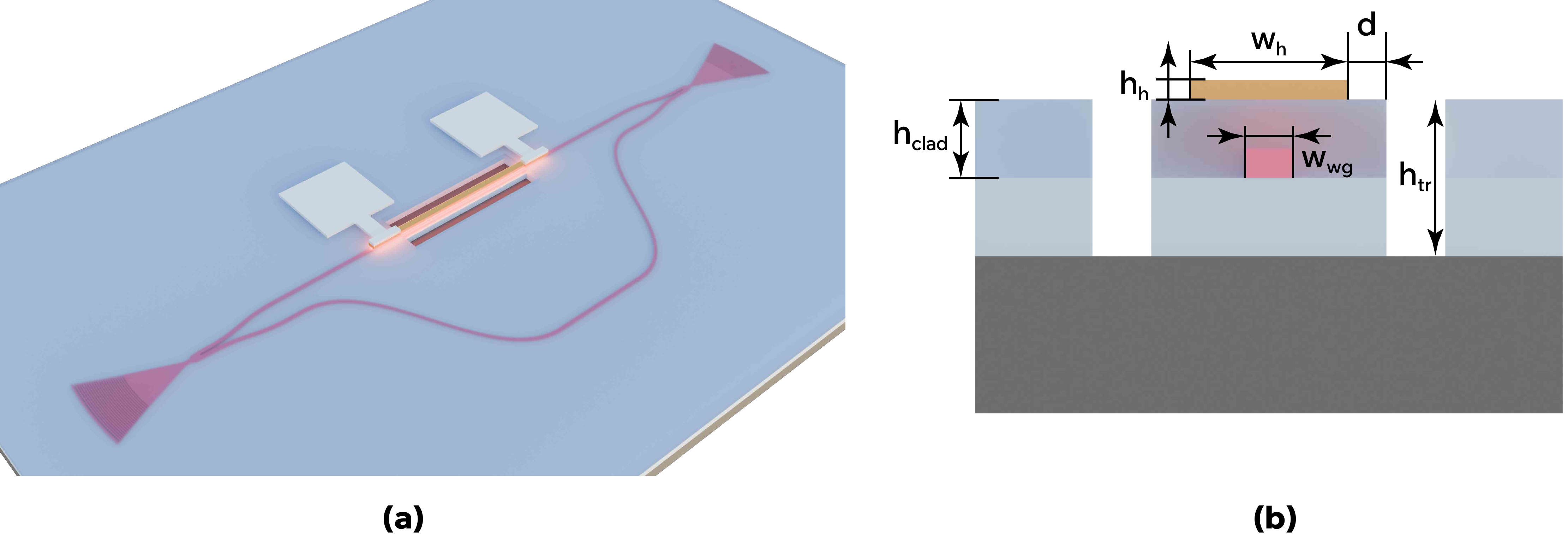}
	\caption{(a) The 3D schematic of the phase shifter structure in unbalanced MZI architecture. (b) Cross-sectional view of the thermo-optic phase shifter.\label{TOPS:Fig3}}
\end{figure}

We designed a phase shifter with a single-strip titanium heater and isolation trenches on a silicon substrate with a thermal SiO$_2$ and a 220 nm-thick stoichiometric silicon nitride. Figure \ref{TOPS:Fig3}(a) shows a 3D schematic of the proposed phase shifter structure. We integrated this phase shifter into an unbalanced MZI with two $1\times2$ Y-branch splitters and a 200 \textmu m arm length difference to measure the thermally induced phase shift. 

In general, the thermally induced phase shift between two spatial modes is described by the following relation \cite{10.1007/s12200-022-00012-9}:
\begin{equation}\label{TOPS:Eq1}
	\Delta\varphi=\dfrac{2\pi L}{\lambda} \left(\dfrac{\mathrm{d}n_\text{eff}}{\mathrm{d}T}\right) \Delta T,
\end{equation}
where $\mathrm{d}n_\text{eff}/\mathrm{d}T$ is the thermo-optic coefficient of the waveguide, $\Delta T$ is the change of temperature.

Because the temperature increase is produced by a resistive heater, expression \eqref{TOPS:Eq1} can be written as:
\begin{equation}\label{TOPS:Eq2}
	\Delta\varphi=\frac{2\pi}{\lambda}\frac{\eta P}{C_{\text{wg}} \rho S}\left(\frac{\mathrm{d}n_{\text{eff}}}{\mathrm{d}T}\right),
\end{equation}
where $\eta$ is the heat transfer coefficient, $P$ is the power consumed by the resistive heater, $C_\text{wg}$ is the heat capacity of the waveguide, $\rho$ is the density of the waveguide material, $S$ is the cross-sectional area of the waveguide.

The electrical power required to shift the phase by $\pi$ is written as follows:
\begin{equation}\label{TOPS:Eq3}
	P_\pi = \frac{\lambda C_{\text{wg}} \rho S}{2 \eta} \left(\frac{\mathrm{d}T}{\mathrm{d}n_{\text{eff}}}\right).
\end{equation}
It should be noted that the performance of the thermo-optic phase shifter is primarily determined by the heat transfer coefficient and is independent of the length of the device. To increase the power efficiency, it is essential to minimize heat leakage into the surrounding environment. When choosing the length of the device, the properties of the material should be taken into account to prevent the heater from transitioning to the liquid phase due to high heating temperatures.

\begin{table}[b] 
	\caption{Physical properties of the materials used in thermo-optic phase shifter design optimization \cite{10.1364/OE.27.010456, 10.3390/photonics8110496, 978-0-8493-0486-6}. \label{TOPS:Table2}}
	\centering
	\begin{tabularx}{1\textwidth}{ >{\centering\arraybackslash}X | >{\centering\arraybackslash}X | >{\centering\arraybackslash}X | >{\centering\arraybackslash}X | >{\centering\arraybackslash}X}
		\hline
		\textbf{Material} & \textbf{Density (g/sm$^\mathbf{3}$)} & \textbf{Specific heat (J/(g$\cdot$K))} & \textbf{Thermal conductivity (W/(m$\cdot$K))} & \textbf{Electrical conductivity (Sm/m)} \\ 
		\hline
		Si           & 2.33 & 0.711 & 148  & 4.3$\times$10$^{-4}$ \\
		SiO$_2$      & 2.20 & 0.709 & 1.38 & 1$\times$10$^{-11}$  \\
		Si$_3$N$_4$  & 3.17 & 0.800 & 30.5 & 1$\times$10$^{-11}$  \\
		Ti           & 4.50 & 0.523 & 21.9 & 0.23$\times$10$^{7}$ \\
		\hline
	\end{tabularx}
\end{table}

To guide the design optimization of a thermo-optic phase shifter, a two-dimensional multiphysics model was developed using the Finite Element Method (FEM). This model integrates electrothermal and optical physics in order to simulate the device's response to applied electrical power. This allows for a systematic evaluation of how geometric parameters affect the required power for a $\pi$ phase shift ($P_\pi$). The simulation involved a two-stage process. First, an electrothermal analysis was performed to determine the steady-state temperature distribution due to Joule heating in the titanium heater. This was accomplished by employing the materials characteristics, which are shown in Table \ref{TOPS:Table2}. A fixed temperature of 293~K was set on the bottom surface of the silicon substrate, while other external boundaries were treated as thermally insulating. Subsequently, the resulting temperature profile was used in optical mode analysis to calculate the temperature-induced change in the effective refractive index of the guided mode ($n_\text{eff}$). The total phase shift over the active device length ($L$) was then determined using the relation: 
\begin{equation}\label{TOPS:Eq4}
	\Delta\varphi=\dfrac{2\pi L}{\lambda} \left(n_\text{eff}(P) - n_\text{eff}(0)\right).
\end{equation}

\begin{figure}[t]
	\centering
	\includegraphics[width=\linewidth]{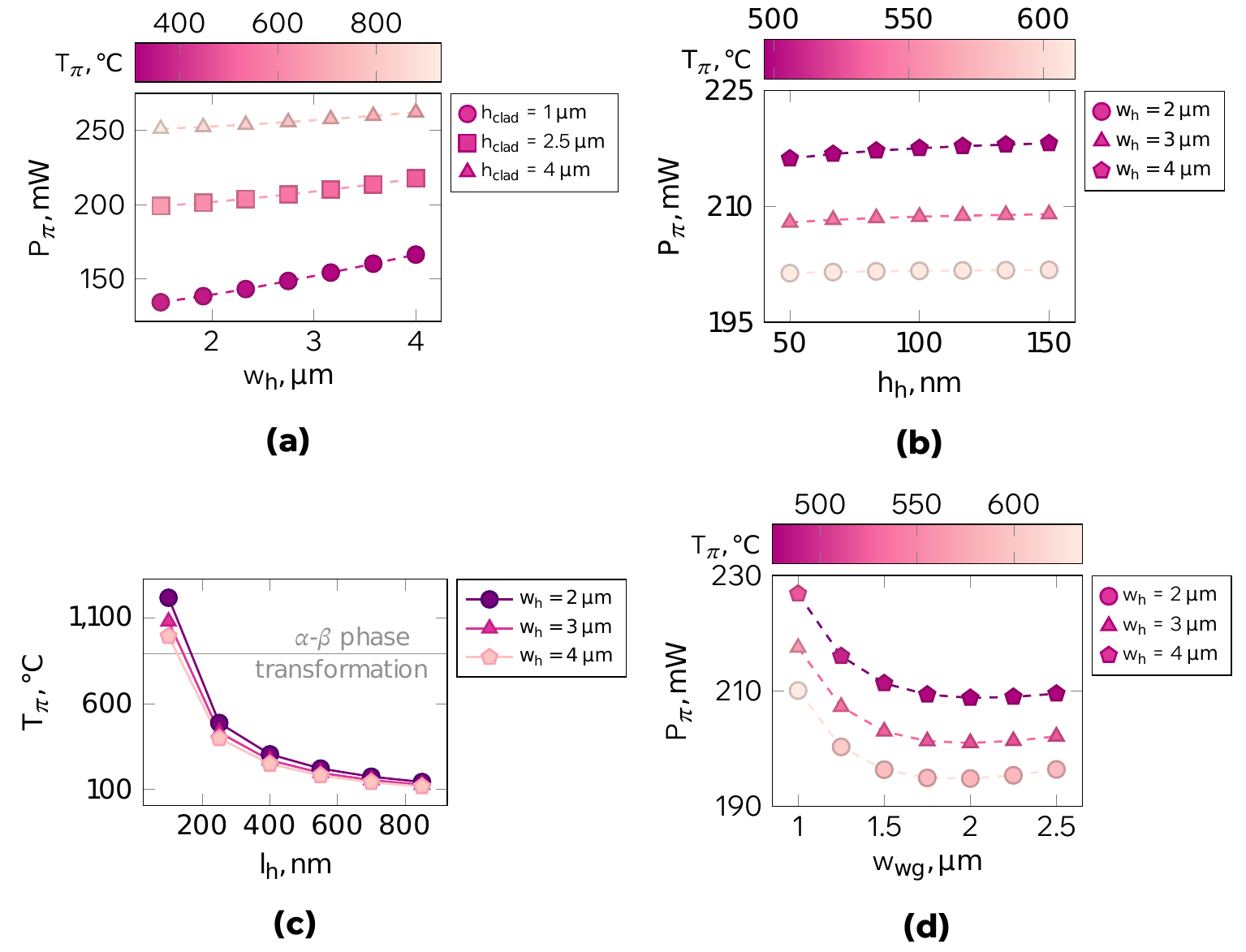}
	\caption{Parametric dependence of phase shifter performance on geometric factors. (a) Power consumption and heater temperature as functions of heater width and cladding thickness. (b) Power consumption and heater temperature as functions of heater width and height. (c) Heater temperature as a function of heater length. (d) Power consumption and heater temperature as functions of heater width and waveguide width. \label{TOPS:Fig4}}
\end{figure} 

A parametric study revealed that several design features significantly influenced $P_\pi$. The effect of the width, thickness, and position of the heater relative to the waveguide, as well as the width of the waveguide and the depth and position of the isolation trenches, on the $P_\pi$ was determined. Results are summarized in Figure \ref{TOPS:Fig4} (a)-(d). The width of the heater has a significant impact on its performance. A narrower heater more efficiently concentrates the heat around the waveguide, leading to reduced power consumption for all geometric configurations considered (Figure \ref{TOPS:Fig4}(a)). The width of 2.0~microns has been chosen as the optimal value, demonstrating the best performance for the current technological capabilities. Similarly, reducing the thickness of the heater can slightly reduce power consumption (Figure \ref{TOPS:Fig4}(b)). However, in order to minimize potential reliability issues associated with electromigration, a thickness of 100~nm was chosen as the optimal value. As additionally shown in Figure \ref{TOPS:Fig4}(a,b), the efficiency of phase shifters depends almost linearly on the thickness of the SiO$_2$ cladding. However, as noted earlier, the proximity of the heater to the waveguide leads to a significant increase in optical loss. In the optimal design, we chose a cladding thickness of 2.5~microns to ensure efficient coupling through diffraction gratings. To prevent thermal damage and change in electrical properties, a minimum heater length of 200~\textmu m was also established to keep operating temperatures below threshold for allotropic $\alpha$ to $\beta$ phase transformation ($\sim$890~$^\circ$C) \cite{10.1063/1.5132739}, as shown in Figure \ref{TOPS:Fig4}(c). 

Additionally, we investigated the effect of the width of the waveguide under the heater on the change in power consumption for a phase shifter design without isolation trenches (Figure \ref{TOPS:Fig4}(d)). As can be seen from equation \eqref{TOPS:Eq3}, the $\pi$-phase shift power consumption is functionally dependent on the cross-sectional area of the waveguide under the heater. Consequently, increasing the waveguide width results in an asymptotic reduction in power consumption; however, the magnitude of this effect diminishes for larger waveguide widths. The change in efficiency is less than 5\%, so, in our current work, we use single-mode waveguides to avoid higher-order modes excitation and increased optical loss. The single strip titanium heater without isolation trenches shows power consumption of 200~mW.

\begin{figure}[b]
	\centering
	\includegraphics[width=\linewidth]{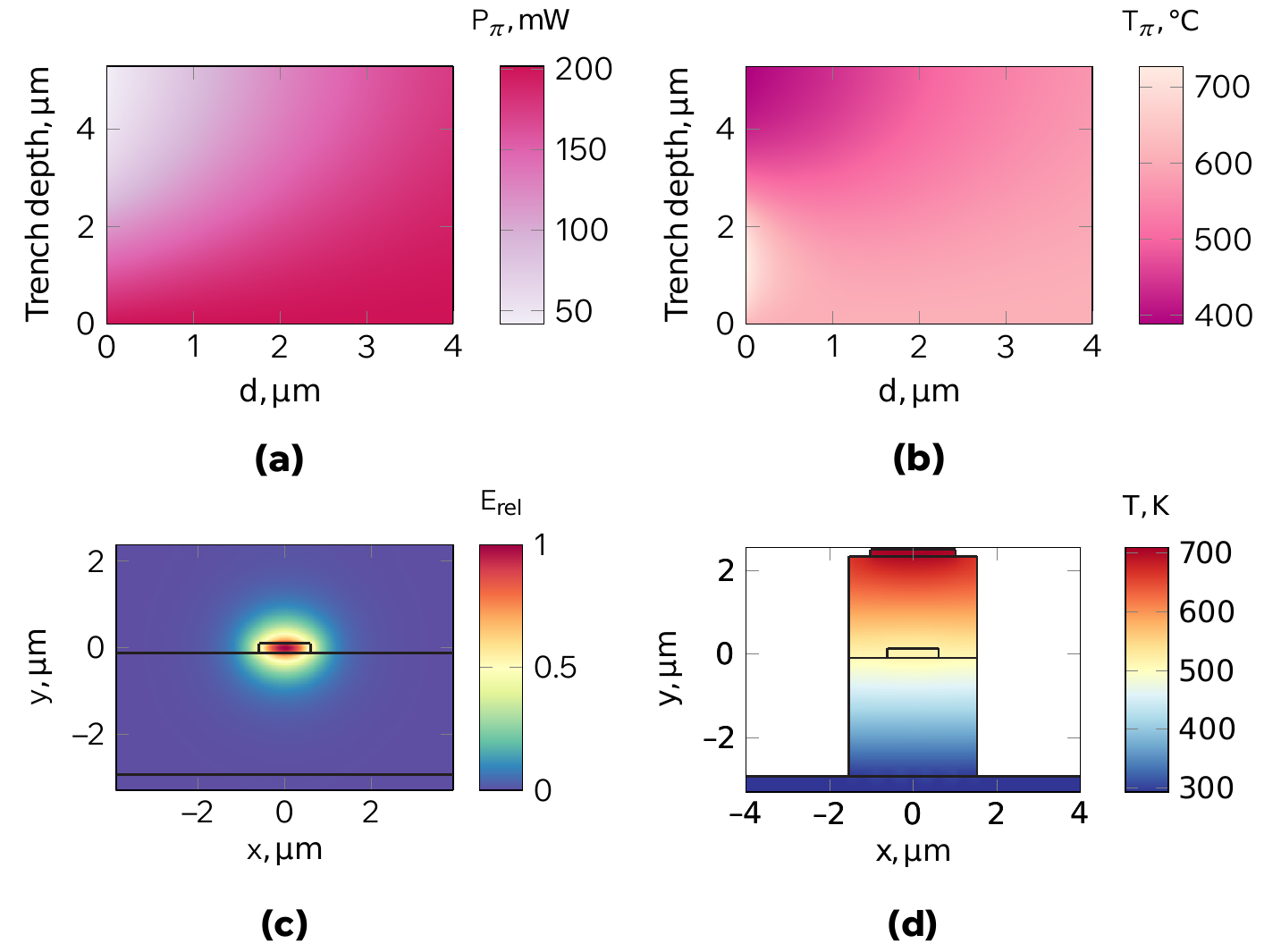}
	\caption{Dependence of phase shifter performance and field distribution on trench geometry. (a-b) Heatmaps of power consumption and corresponding heater temperature at the $\pi$-phase shift, both as functions of trench distance from the heater and trench depth. (c) Visualization of the electromagnetic field distribution within the phase shifter cross-section. (d) Temperature distributions in the optimized phase shifter cross-section.\label{TOPS:Fig5}}
\end{figure}   

As discussed in the introduction, the most effective solutions for reducing power consumption in phase shifters are still the use of air insulation trenches and folded waveguides that pass under the phase shift a certain number of times. However, the design of a phase shifter using folded waveguides presents challenges for telecommunication wavelengths. The gap between the waveguides needs to be significantly larger than the $\lambda$ in order to avoid excessive coupling, which leads to an increase in the final device’s footprint. To further enhance thermal efficiency, air trenches were incorporated adjacent to the heater to act as thermal barriers. The model indicated a strong dependence of $P_\pi$ on the position of the trenches $d$, with closer proximity improving thermal confinement, providing a power consumption of 42~mW (Figure \ref{TOPS:Fig5}(a)). Our simulations indicated that a trench depth of at least 4~\textmu m might be necessary to provide a substantial reduction in lateral heat loss toward the SiO$_2$ (Figure \ref{TOPS:Fig5}(a, b)). We have chosen the distance between the heater and the air trench to be 0.5~microns as optimal to prevent possible damage to the heater during the manufacturing process. The final optimized phase shifter design produced a $\pi$ phase shift in the model at an applied electrical power of 64~mW for a compact active length of 200~\textmu m. Under these simulated conditions, the peak temperature of the heater was approximately 710~K, and the temperature of the waveguide core was around 508~K. The fundamental TE mode at 1550~nm in the strip Si$_3$N$_4$ waveguide is shown in Figure \ref{TOPS:Fig5}(c). In Figure \ref{TOPS:Fig5}(d), the temperature distribution is shown for the optimized design of a single strip phase shifter with isolation trenches at $P_\pi$.

\section{Results}

\subsection{Device fabrication \label{TOPS:Sec31}}

\begin{figure}[h!]
	\centering
	\includegraphics[width=\linewidth]{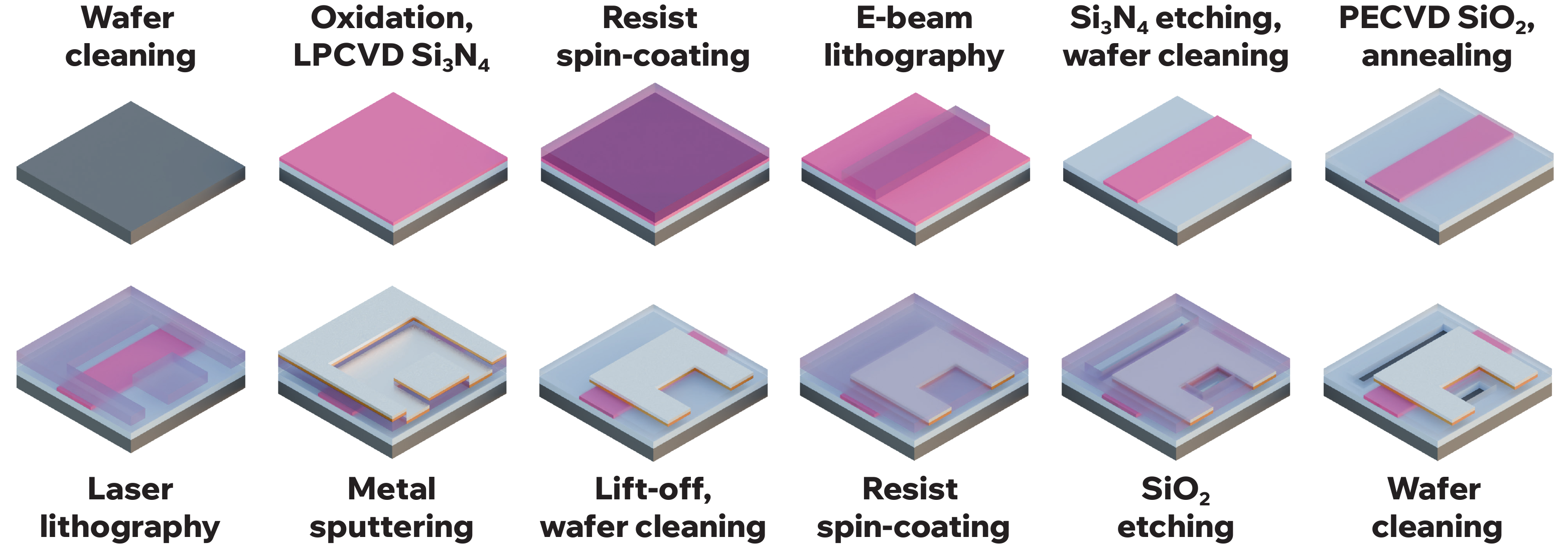}
	\caption{The fabrication process of thermally-tunable, low-loss silicon nitride photonic integrated  circuits.\label{TOPS:Fig6}}
\end{figure}

The proposed thermo-optic phase shifters were fabricated at the BMSTU Nanofabrication Facility (Shukhov Labs) as part of our silicon nitride integrated photonics platform \cite{10.1364/OE.477458, 10.3390/mi16121401}. The fabrication process is outlined schematically in Figure \ref{TOPS:Fig6}. After an initial cleaning of 4" silicon wafers, we formed a buried oxide (BOX) layer via wet thermal oxidation, followed by deposition of a 220~nm-thick stoichiometric silicon nitride film by low-pressure chemical vapor deposition (LPCVD). Each process step was followed by wafer cleaning and annealing at 1100~\textdegree C. We then used proprietary electron-beam lithography (EBL) and inductively coupled plasma  reactive ion etching (ICP-RIE) to create waveguide cores with low sidewall roughness \cite{10.3390/mi16121401}. Next, a SiO$_2$ cladding was deposited by plasma-enhanced chemical vapor deposition (PECVD) to encapsulate the waveguides. The heaters were patterned using photolithography and a metal lift-off process, including  magnetron sputtering of titanium. Finally, deep etching of the SiO$_2$ was performed to define the isolation trenches.

\begin{figure}[t]
	\centering
	\includegraphics[width=.965\linewidth]{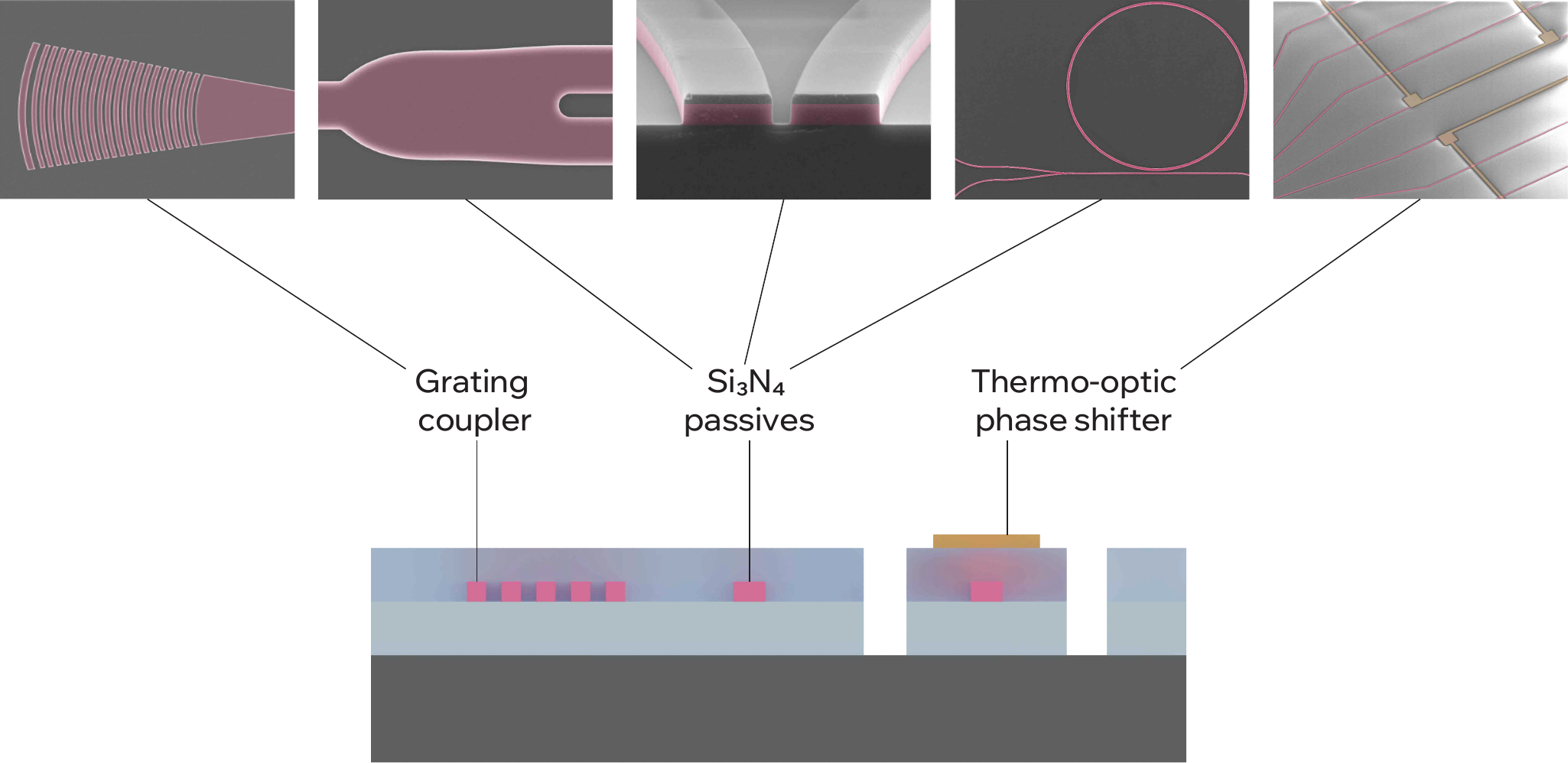}
	\caption{The optical microscope image of the fabricated MZI structure. Insets: the SEM images of waveguide, grating coupler, y-splitter and thermo-optic phase shifter. \label{TOPS:Fig7}}
\end{figure}

Figure \ref{TOPS:Fig7} shows an optical micrograph of one of the fabricated MZI structures and input/output grating couplers designed for TE polarization. The insets show scanning electron microscope (SEM) images of cross-section of the Si$_3$N$_4$ waveguide with SiO$_2$ cladding, grating coupler, y-splitter and thermo-optic phase shifter.

\subsection{Device characterization: analysis of the performance of thermo-optic phase shifters}

\begin{figure}[h!]
	\centering
	\includegraphics[width=0.69\linewidth]{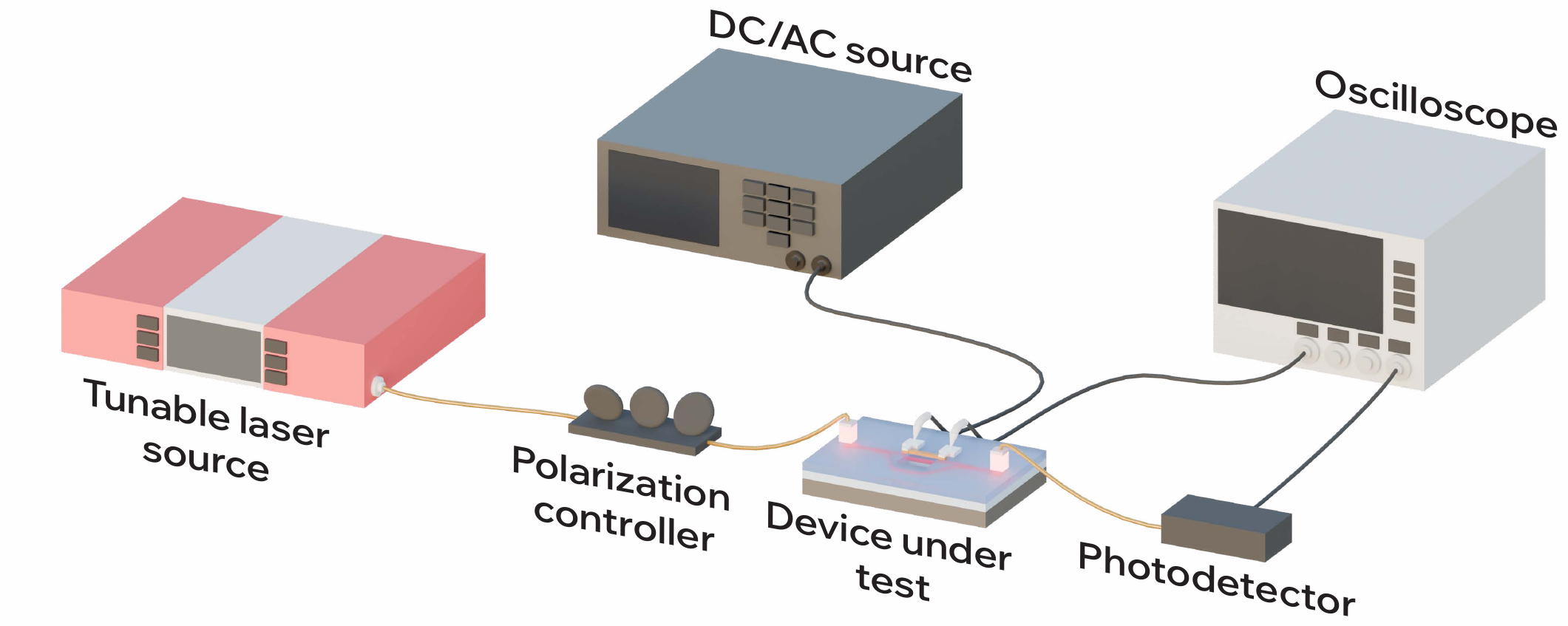}
	\caption{Measurement setup for DC and AC characterization.\label{TOPS:Fig8}}
\end{figure}

For performance characterization of the fabricated thermo-optic phase shifters, we used the DC and AC measurement setup illustrated in Figure \ref{TOPS:Fig8}.

We determined the phase shift from the measured transmission spectra of the fabricated unbalanced MZI structures according to \cite{10.1364/OL.436288}:

\begin{equation}\label{TOPS:Eq5}
	\Delta\varphi=\dfrac{\left| \lambda(V_0) - \lambda(V) \right|}{\text{FSR}}
\end{equation}

\begin{figure}[t]
	\includegraphics[width=0.98\linewidth]{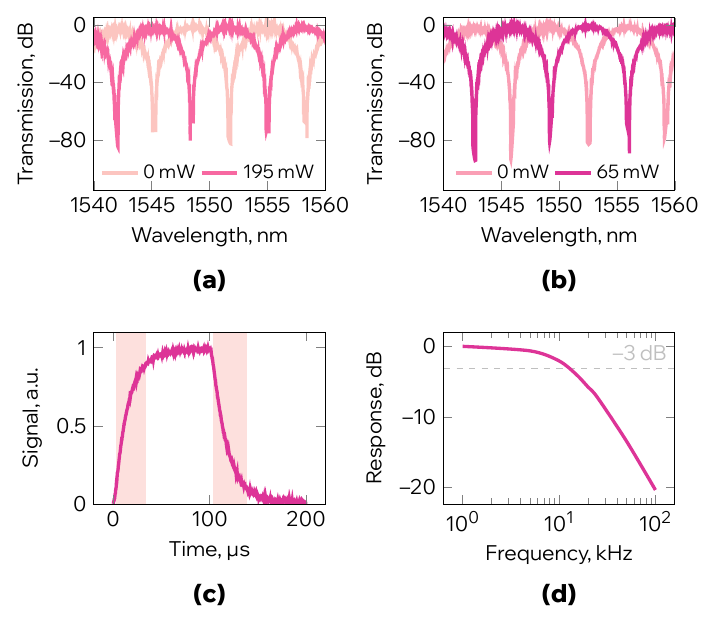}
	\caption{Experimental characterization of thermo-optic phase shifters. (a-b) Transmission spectra at $P=0$ and $P=P_\pi$ for devices without and with trenches, respectively. (c) Temporal optical response to square voltage pulses, with measured rise time $T_\text{rise}=31.6$~\textmu s and fall time $T_\text{fall}=35.4$~\textmu s. (d) Optical response versus modulation frequency showing the 12 kHz bandwidth of the thermo-optic phase shifter, determined at the $-3$ dB cutoff point.\label{TOPS:Fig9}}
\end{figure}

Figure \ref{TOPS:Fig9}(a,b) shows the normalized transmission spectra for fabricated devices, yielding an extinction ratio of $\sim 80$~dB and $\pi$ phase shift power consumption of 65~and 195~mW for MZI with and without isolation trenches, respectively.

The dynamic performance of the device is further characterized by driving the heater with a square waveform electrical signal when the interferometer is in half the transmission \cite{10.3390/photonics11070603}. The waveform of the output optical signal is measured via a photodetector followed by an oscilloscope. The 10-90\% rise and fall times were measured to be $32$ and $35$~\textmu s, respectively, for the phase shifters with air trenches (Figure \ref{TOPS:Fig9} (c)). To evaluate the bandwidth of the thermal phase shifter, the frequency response of the Mach-Zehnder modulator was measured. A sinusoidal voltage was applied to the heater while monitoring the modulated optical output, revealing a $-3$~dB cutoff frequency of $12$~kHz (Figure \ref{TOPS:Fig9} (d)). 

\subsection{Device characterization: analysis of propagation loss}

\begin{figure}[t]
	\centering
	\includegraphics[width=0.98\linewidth]{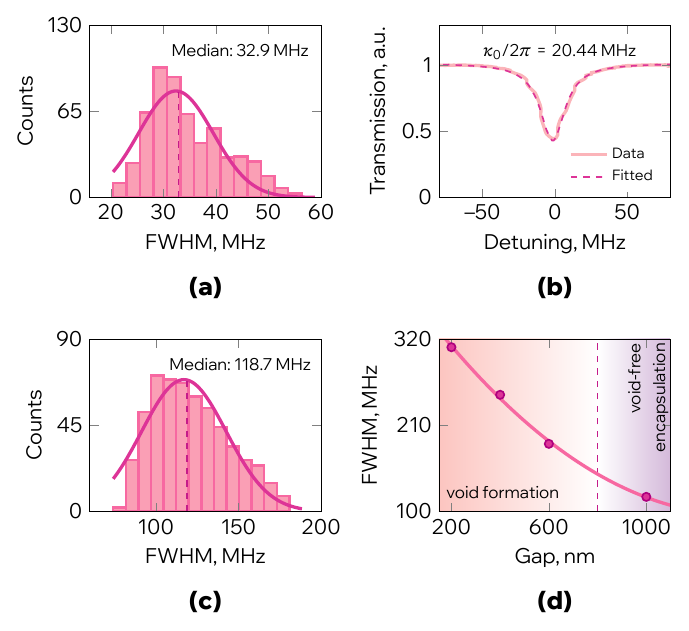}
	\caption{Intrinsic loss characterization of Si$_3$N$_4$ microring resonators. (a) Histogram of intrinsic linewidths $k_0/2\pi$ with most probable 32.9~MHz for multimode waveguide. (b) representative resonance at 1565.0~nm with fitted lineshape yielding $k_0/2\pi = 20.44$~MHz. (c) Histogram of intrinsic linewidths $k_0/2\pi$ with most probable 118.7~MHz for single-mode waveguide. (d) FWHM versus coupling gap showing further loss reduction at larger gaps due to mitigated void-related loss in the coupler region. \label{TOPS:Fig10}}
\end{figure}  

We investigated the impact of the fabrication process of phase shifters on propagation loss by measuring the intrinsic resonance linewidths ($k_0/2\pi$) of microring resonators with transverse electric polarization using frequency-calibrated laser spectroscopy \cite{10.1364/PRJ.521172}. The histogram on Figure \ref{TOPS:Fig10}(a) shows the most probable value of $k_0/2\pi = 32.9$~MHz, corresponding to most probable intrinsic quality factor ($Q_0$) of $5.9\times10^6$ and propagation loss of 0.058~dB/cm for microresonators of 10~GHz free spectral range (FSR) and 2400~nm width. A minimum intrinsic linewidth of $k_0/2\pi = 20.44$~MHz is observed at 1565.0~nm (Figure \ref{TOPS:Fig10}(b)), corresponding to $Q_0 = 9.6\times10^6$ and $\alpha=0.033$~dB/cm. Subsequently, we analyzed the dependence of propagation loss on the waveguide width and coupling gap. Propagation loss is mainly caused by the interaction of light with the roughness of the waveguide’s surface, especially at the sidewalls. One of the most effective ways to reduce loss is by increasing the width of the waveguide. This reduces the interaction between optical mode and the surface roughness, leading to a decrease in the intrinsic linewidth from 118.7~MHz for single-mode waveguides (Figure \ref{TOPS:Fig10}(c)). The widening of the gap between the waveguide and microring can significantly contribute to a more efficient reduction of loss, as illustrated in the Figure \ref{TOPS:Fig10}(d). The main reason involved here is the presence of additional loss in the coupling region due to the void formation, which is a known challenge for SiO$_2$ cladding deposition (PECVD or LPCVD) in the subtractive silicon nitride waveguide fabrication \cite{10.1364/OPTICA.3.000020}. 

\section{Conclusion}

In summary, we have demonstrated a compact and power-efficient silicon nitride thermo-optic phase shifters for the C-band. The proposed phase shifters were integrated into MZI structures, showing low $P_\pi$ of 65~mW and 195~mW for device with and without isolation trenches, respectively, extinction ratio over 80~dB, and the measured bandwidth over 12 kHz. The phase shifters were fabricated on 100-mm diameter Si wafers as part of our low-loss integrated photonics platform, maintaining microring resonators with an intrinsic quality factor of 5.9 million and propagation loss below 0.058~dB/cm. Our approach reveals its potential in R\&D and scalable producton of reconfigurable photonic circuits for promising future applications in quantum technologies, high-performance computing, surveying and navigation, and medical treatment and diagnosis. 

\begin{backmatter}
	\bmsection{Funding}
	This research received no external funding.
	
	\bmsection{Acknowledgment}
	Technology was developed, and samples were fabricated and measured at Quantum Park (BMSTU Nanofabrication Facility, Shukhov Labs, FMNS REC, ID 74300).
	
	\bmsection{Disclosures}
	The authors declare no conflicts of interest.
	
	\bmsection{Data Availability Statement}
	The data that supports the findings of this study are available from the corresponding author upon reasonable request.
\end{backmatter}


\bibliography{TOPS-bibliography}

\end{document}